# Angular momentum of the physical electron

## A. M. Stewart

Department of Theoretical Physics,
Research School of Physics and Engineering,
The Australian National University,
Canberra, ACT 0200, Australia.

**Abstract** The angular momentum of the physical electron, modeled as a Dirac fermion coupled to the electromagnetic field, is found to be $\hbar/2$, the same as that of a bare Dirac fermion and independent of the sign and magnitude of the electric charge. The contribution of the electromagnetic field to the angular momentum is zero.

The coupled Maxwell-Dirac equations

$$(i\gamma^\nu \partial_\nu - \kappa)\psi = \frac{e}{\hbar c} A_\nu \gamma^\nu \psi \quad \text{and} \quad \partial_\sigma F^{\sigma\mu} = 4\, e\bar{\psi}\gamma^\mu \psi = 4\, j^\mu/c \quad , \quad (1)$$

where $\psi$ is a fermion field operator and $\kappa = mc/\hbar$, that arise from the Lagrangian density $\mathcal{L}$ of quantum electrodynamics

$$\mathcal{L}/\hbar c = \frac{i}{2}\bar{\psi}\gamma^\sigma \partial_\sigma \psi - \frac{i}{2}(\partial_\sigma \bar{\psi})\gamma^\sigma \psi - \kappa \bar{\psi}\psi - \frac{e}{\hbar c} A_\sigma \bar{\psi}\gamma^\sigma \psi - \frac{1}{16\,\hbar c} F^{\alpha\beta} F_{\alpha\beta} \quad (2)$$

imply that the physical electron may be modeled as a composite particle consisting of the bound state of a Dirac fermion with bare mass *m* and charge *e* and the electromagnetic field [1, 2]. The electromagnetic field *F* and potential *A* are each composed of two parts: the part due to external sources and the part due to the electron's own (self) field. For the case of an isolated electron, which we consider, the external parts are set to zero.

Early attempts to calculate physical properties from (2) by perturbation theory led to divergences [3], and more recent attempts to solve (1-2) as classical equations [4] or by variational methods [5, 6] have met with only limited success. In this paper we derive an operator relation for the angular momentum of the physical electron and demonstrate that it has the same spin as a bare Dirac fermion, namely $\hbar/2$.

The symmetric energy-momentum tensor $T^\mu$ that corresponds to the above Lagrangian density is obtained by the usual method [1, 2, 7] to be

$$T^{\mu\nu} = c\hbar \frac{i}{4}(\bar{\psi}\gamma^\mu D^\nu \psi - D^\nu {}^* \bar{\psi}\gamma^\mu \psi + \bar{\psi}\gamma^\nu D^\mu \psi - D^\mu {}^* \bar{\psi}\gamma^\nu \psi) + \frac{1}{4}(\frac{\eta^{\mu\nu}}{4} F_{\alpha\beta} F^{\alpha\beta} + F^{\mu\sigma} F_\sigma{}^\nu) \quad (3)$$

where $D^\mu = \partial^\mu + \frac{ie}{\hbar c} A^\mu$ is the gauge covariant derivative. The electromagnetic part of the tensor has been made symmetric and gauge invariant by the method of Belinfante [2, 8]. The Hamiltonian of the system is the integral of $T^{00}$ over all space. The physical electron is described by the ground eigenstate of the QED Hamiltonian [1].

From (3), the angular momentum operator $\boldsymbol{J}$ of the coupled system is found to be

$$\boldsymbol{J} = -i\hbar \int d^3x\, \boldsymbol{x} \times \psi^\dagger \nabla \psi - \frac{e}{c} \int d^3x\, \boldsymbol{x} \times \psi^\dagger \boldsymbol{A}\psi + \frac{\hbar}{2} \int d^3x\, \psi^\dagger \boldsymbol{\Sigma} \psi + \frac{1}{4\pi c} \int d^3x\, \boldsymbol{x} \times (\boldsymbol{E} \times \boldsymbol{B}) \quad (4)$$

where the four-component spin operator $\Sigma$ is

$$\Sigma = \begin{pmatrix} \sigma & 0 \\ 0 & \sigma \end{pmatrix}, \quad (5)$$

$\sigma$ being the three Pauli matrices [2]. The first two terms of (4) are the gauge invariant orbital angular momentum of the fermion, the third term is the spin and the last term describes the angular momentum of the electromagnetic field. The calculation of the Dirac parts of (4) from (3) is given by Ohanian [9]. The expression for $\boldsymbol{J}$ is gauge invariant. Considerations of the angular momentum of composite systems have been of interest recently in attempts to resolve the spin crisis of hadron physics [1, 10, 11].

The last term of (4), the angular momentum of the electromagnetic field, has been examined recently [12-15]. It has been shown that it can be separated into two parts a radiative part and a bound part. The radiative part may in turn be decomposed into three terms, a spin term, an orbital term and a boundary term. The spin term and orbital term have been shown to give the standard expressions for paraxial waves [15]. The spin term and the boundary term have been shown to resolve the paradox involving the angular momentum of planes waves of electromagnetic radiation [13]. However, because a quantum system in its groundstate does not radiate, only the bound part needs to be considered in the context of the present paper. The only relevant part of the last term of (4) therefore is the bound part [14, 16], the part associated with the electron's self-field, which is

$$\boldsymbol{J}_b = \frac{1}{c} \int d^3x\, \rho(\boldsymbol{x}) \boldsymbol{x} \times \boldsymbol{A}_t(\boldsymbol{x}) \quad (6)$$

where $\boldsymbol{A}_t$ is the vector potential in the transverse or Coulomb gauge (div$\boldsymbol{A}_t = 0$) [17, 18] and $\rho$ is the charge density, in the present case $e\psi^\dagger(\boldsymbol{x})\psi(\boldsymbol{x})$. The transverse vector potential $\boldsymbol{A}_t$ can be expressed explicitly in terms of the instantaneous magnetic field [17, 19]. It will be seen that the precise form need not concern us.

It is useful to work in the Coulomb gauge so that $\boldsymbol{A}$ becomes $\boldsymbol{A}_t$ in (4) and we find that (6), which replaces the last term of (4), cancels the second term of (4), so that (4) simplifies to give

$$\boldsymbol{J} = -i\hbar \int d^3x\, \boldsymbol{x} \times \psi^\dagger \nabla \psi + \frac{\hbar}{2} \int d^3x\, \psi^\dagger \boldsymbol{\Sigma} \psi \quad . \quad (7)$$

Although (7) gives the angular momentum operator of the full interacting system it has the same form as for a non-interacting Dirac fermion and accordingly has the same eigenvalues. Therefore, the spin of the physical electron, like that of the Dirac fermion, is predicted to have a magnitude of $\hbar/2$, independent of the size of the electronic charge. Although the electromagnetic field has an angular momentum given by the last term of (4) [20], it is exactly cancelled by the term in (4) that exhibits the vector potential explicitly. Since (4) is gauge invariant the argument holds in any gauge. Equation (7) obeys the standard quantum commutation relations for angular momentum.

By similar but simpler arguments, it can be shown that the linear momentum operator of the interacting system is the same as that of the bare fermion alone; the electromagnetic field makes no contribution.

From the experimental fact that the spin of the physical electron is $\hbar/2$, it is clear that there are only two possibilities for the magnitude of the angular momentum associated with the electromagnetic field. It could be of magnitude $\hbar$; in this case the fermion spin would have to couple antiparallel to the angular momentum associated with the electromagnetic field to give the experimental value of spin $\hbar/2$ for the physical electron. The second possibility is that the angular momentum associated with the electromagnetic field is zero and the spin of the system is therefore just that of the fermion alone: $\hbar/2$. It has been shown in this paper, by direct calculation of the operator for the angular momentum, that the second possibility is correct and the first is incorrect.